\documentclass[prl,showpacs,twocolumn,superscriptaddress]{revtex4-1}
\usepackage[utf8]{inputenc}
\usepackage[american,british]{babel}
\usepackage[T1]{fontenc}
\usepackage[pdftex]{graphicx}  
\usepackage{graphicx, xcolor}
\usepackage{dcolumn}
\usepackage{bm}
\usepackage{amsmath,amsthm,amssymb}
\usepackage{amsmath}
\usepackage{color}
\usepackage{verbatim}
\usepackage{ulem}

\definecolor{darkGreen}{RGB}{0,110,0}
\definecolor{darkBlue}{RGB}{0,0,130}
\usepackage[colorlinks,citecolor=darkGreen,linkcolor=darkBlue,%
urlcolor=blue,hyperindex]{hyperref}

\newcommand{\cd}[1]{c^{\dagger}_{#1}}

\begin{document}
\title{Majorana Quasi-Particles Protected by $\mathbb{Z}_2$ Angular Momentum Conservation}
\author{F. Iemini}
\address{Abdus Salam International Center for Theoretical Physics, Strada Costiera 11, Trieste, Italy}
\author{L. Mazza}
\address{Departement de Physique, Ecole Normale Superieure / PSL Research University,
CNRS, 24 rue Lhomond, F-75005 Paris, France}
\author{L. Fallani}
\address{Department of Physics and Astronomy, University of Florence, I-50019 Sesto Fiorentino, Italy}
\address{LENS European Laboratory for Nonlinear Spectroscopy, I-50019 Sesto Fiorentino, Italy}
\author{P. Zoller}
\address{Institute for Theoretical Physics, University of Innsbruck, A-6020 Innsbruck, Austria}
\address{Institute for Quantum Optics and Quantum Information of the Austrian Academy of Sciences, A-6020 Innsbruck, Austria}
\author{R. Fazio}
\address{Abdus Salam International Center for Theoretical Physics, Strada Costiera 11, Trieste, Italy}
\address{NEST, Scuola Normale Superiore and Istituto Nanoscienze-CNR, I-56126 Pisa, Italy}
\author{M. Dalmonte}
\address{Abdus Salam International Center for Theoretical Physics, Strada Costiera 11, Trieste, Italy}
\date{\today}

\begin{abstract}
We show how angular momentum conservation can stabilise a symmetry-protected quasi-topological phase of matter supporting Majorana quasi-particles as edge modes in one-dimensional cold atom gases. We investigate a number-conserving four-species Hubbard model in the presence of spin-orbit coupling. The latter reduces the global spin symmetry to an angular momentum parity symmetry, which provides an extremely robust protection mechanism that does not rely on any coupling to additional reservoirs. The emergence of Majorana edge modes is elucidated using field theory techniques, and corroborated by density-matrix-renormalization-group simulations. Our results pave the way toward the observation of Majorana edge modes with alkaline-earth-like fermions in optical lattices, where all basic ingredients for our recipe - spin-orbit coupling and strong inter-orbital interactions - have been experimentally realized over the last two years.
\end{abstract} 

\maketitle

\paragraph{Introduction. -- } The past two decades have witnessed an impressive progress in understanding how to harness quantum systems supporting topological order, one of the ultimate goals being the observation of quasi-particles with non-Abelian statistics -- non-Abelian anyons~\cite{nayak2008,wilczek2009,alicea2012,beenakker2013,goldman2016}. A pivotal role in this search has been the formulation of a model for one-dimensional (1D) p-wave superconductors~\cite{kitaev2001}, that supports a symmetry-protected topological phase with Majorana quasi-particles (MQPs) as edge modes. The key element for the stability of such edge modes is the presence of a $\mathbb{Z}_2$ parity symmetry. At the mean-field level, this can be realized via proximity-induced superconductivity in solid-state settings~\cite{Kouwenhoven,Deng,Das,Rokhinson,nadj,marcus}, or via coupling to molecular Bose-Einsten condensates in cold atoms~\cite{liang2011}. Remarkably, it is possible to stabilize MQPs even taking fully into account quantum fluctuations by considering canonical settings~\cite{cheng2011,fidkowski2011,sau2011,Ortiz_2014,Chen_2016}, where the parity symmetry emerges via, e.g., engineered pair-tunneling between pairs of wires~\cite{Kraus2013,Iemini2015,Lang2015}. However, it is an open challenge to understand whether, in these number-conserving setups, there exist {\it fundamental} microscopic symmetries that can serve as a pristine mechanism for the realization of MQPs, that is genuinely distinct from reservoir-induced superconductivity.

\begin{figure}
\includegraphics[width=0.95\columnwidth]{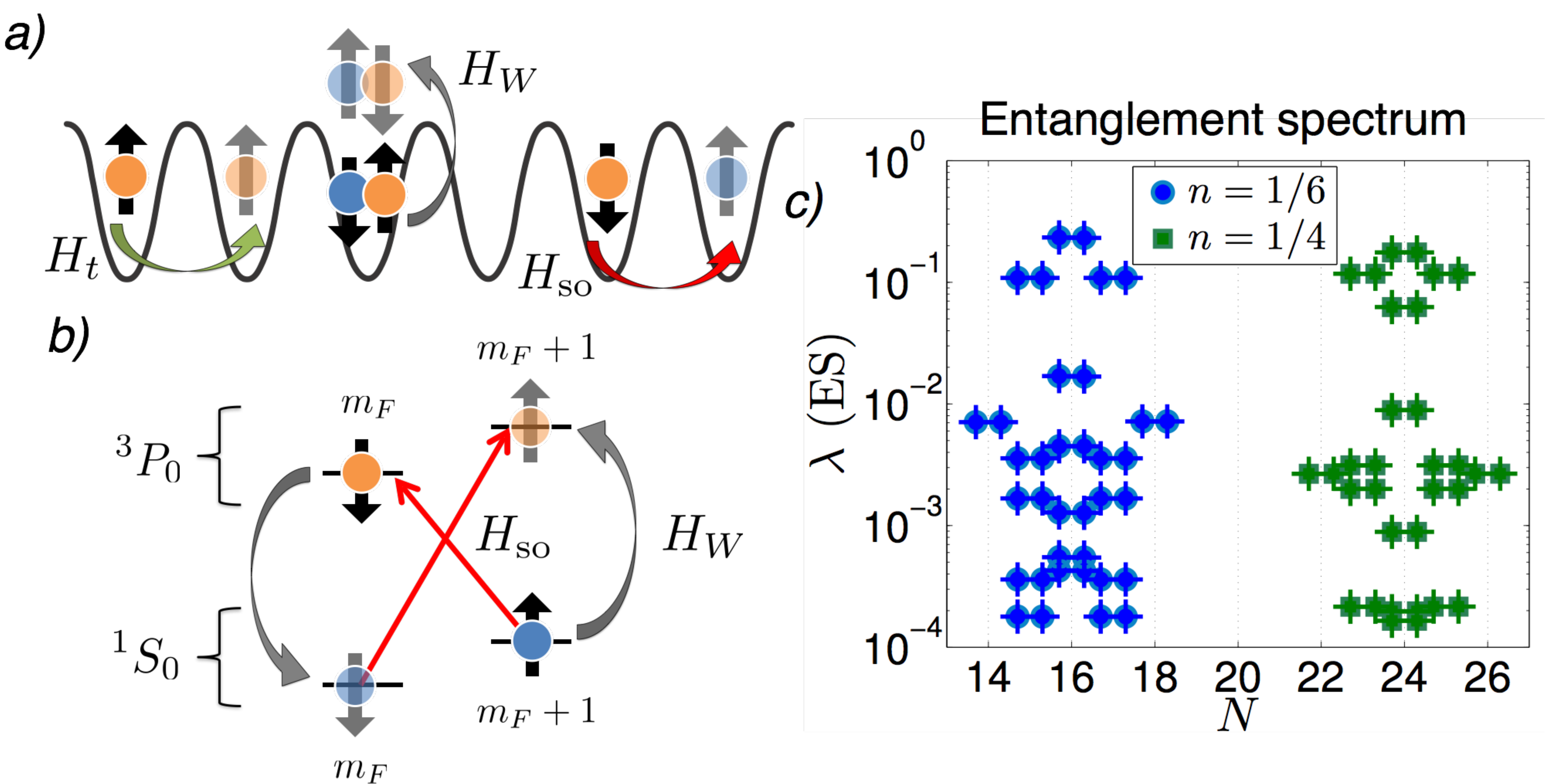}
\centering
\caption{ Schematics of the orbital Hubbard model in the presence of spin-orbit coupling as realized with alkaline-earth-like atoms. 
{\it a-b)} The model we consider in Eq.~\eqref{H0} describes tunneling ($H_t$), spin-orbit-coupling ($H_{\text{so}}$), and spin-exchange processes ($H_W$). In cold atom settings, the spin degree of freedom is represented by different Zeeman states with nuclear spin $m_F, m_F+1$, while the orbital degree of freedom is encoded in different electronic states, $^1$S$_0$ and $^3$P$_0$. In these systems, $H_W$ and $H_{\text{so}}$ are described by the grey and red arrows, respectively. {\it c)} In the quasi-topological phase of the model, the entanglement spectrum displays a characteristic two-fold degeneracy: eigenvalues of the reduced density matrix with the same number of particles come in pairs with opposite parities (see text). }
\label{fig:scheme}
\end{figure} 

Here, we show how angular momentum conservation enables the realization of a symmetry-protected quasi-topological phase supporting MQPs in one-dimensional number-conserving systems~\cite{FootNote1}. In particular, we show how a combination of spin-exchange interactions and {\it crossed} spin-orbit couplings in orbital Hubbard models (see Fig.~\ref{fig:scheme}a-b) naturally gives rise to a $\mathbb{Z}_2$ spin symmetry. This symmetry serves as the enabling tool to realize MQPs, and, as we discuss below, its robustness is guaranteed by the fact that all terms breaking it are not present in the microscopic dynamics, as they would violate angular momentum conservation. Remarkably, these models find direct and natural realization using Alkaline-earth-like atoms (AEAs) in optical lattices~\cite{cazalilla2009,gorshkov2010,cazalilla2014,stellmer2009,desalvo2010,sugawa2011,swallows2011,mancini2015,hofrichter2016}: in these settings, both spin-exchange interactions~\cite{cappellini2014,scazza2014} and spin-orbit couplings~\cite{wall2016,livi2016,kolkowitz2016} have already been demonstrated, providing an ideal setting to realize MQPs using state-of-the-art experimental platforms within the paradigm described in the present work.

\paragraph{Model Hamiltonian. -}
Our starting point is a one-dimensional Hubbard model describing four fermionic species, with annihilation operators $c_{j,\alpha, p}$, with $j\in[1, L]$ a site index, $L$ the length of the system, $\alpha \in [\uparrow, \downarrow]$ describing a pseudo-spin encoded in a pair of Zeeman states $m_F, m_{F}+1$ (depicted in Fig.~\ref{fig:scheme}a-b as arrows), and $ p \in [-1,1]$ describing orbital degrees of freedom, encoded in the electronic state ground ($^1$S$_0$, blue) and meta-stable ($^3$P$_0$, orange) states. The system Hamiltonian reads:
\begin{eqnarray}\label{H0}
 H = 
\sum_j (H_{t,j}  + H_{U,j} + H_{W,j}+ H_{{\mathrm{so},j}});
\end{eqnarray}
(in the following we also use the notation $ H_x = \sum_j H_{x,j}$).
The first two terms represents tunneling along the wire, ${H}_{t,j}=-\sum_{\alpha, p} t(c^\dagger_{j, \alpha, p}c_{j+1, \alpha, p}+\text{h.c.})$, and diagonal interactions, 
 ${H}_{U,j} = \sum_{p} U_{p}  n_{j,\uparrow,p}  n_{j,\downarrow,p} +  U\sum_{\alpha, \beta}  n_{j,\alpha,-1}  n_{j,\beta,1}$. The third term, visualized by grey arrows in Fig.~\ref{fig:scheme}, describes spin-exchange interactions~\cite{FootNote2}:
\begin{equation}
{H}_{W,j} = W(c^\dagger_{j, \uparrow, -1} c^\dagger_{j, \downarrow, 1} 
c_{j, \downarrow, -1} c_{j, \uparrow, 1} + \text{h.c.}).
\end{equation}
The last term describes a generalized spin-orbit coupling:
\begin{eqnarray}\label{HSO}
{H}_{\rm{so},j} &=& \sum_{p}\left\{(\alpha_R+b) \cd{j,\uparrow,p}c_{j+1,\downarrow,-p}+\right.\nonumber \\
&+& \left.(b-\alpha_R) \cd{j+1,\uparrow,p}c_{j,\downarrow,-p}  + \text{h.c.}\right\},
\end{eqnarray}
where $\alpha_R$ denotes the Rashba velocity, and  the $b$ term may be seen as momentum-dependent Zeeman field~\cite{Budich_2013,Budich_2015, FootNote3}. 

In microscopic implementations, the last two terms in $ H$ are embodied by strong inter-orbital spin-exchange interactions (grey arrows)~\cite{cappellini2014,scazza2014}, and by the possibility of engineering {\it crossed} spin-orbit couplings (red arrows) via clock lasers~\cite{wall2016,livi2016,kolkowitz2016}. The combination of these two ingredients breaks explicitly the global spin-symmetry from $SU(2)\times SU(2)$ down to $\mathbb{Z}_2$ -- namely, the number of states in each pair of states coupled by spin-orbit coupling is conserved {\it modulo} 2, due to the presence of the spin-exchange interactions. Indeed, while for $\alpha_R=b=0$ the Hamiltonian has a $SU(2)\times SU(2)$ spin symmetry~\cite{cazalilla2009,gorshkov2010}, for generic values of $\alpha_R ,b\neq0$, the spin symmetry is reduced to $\mathbb{Z}_2$, whose correspondent conserved charge is the mutual parity between the two subsets $[(\uparrow, 1), (\downarrow, -1)]$ and $[(\uparrow, -1), (\downarrow, 1)]$ connected by the spin exchange interaction ${H}_W$, i.e., $P_{m} = \mod_2 [(\sum_{j} (n_{j,\uparrow, 1} + n_{j,\downarrow, -1}) - (n_{j,\uparrow, -1} + n_{j,\downarrow, 1}))/2]$. The robustness of this emergent parity symmetry stems from from angular momentum conservation: this symmetry may only be broken in the presence of terms such as, e.g., $c^\dagger_{j\uparrow,-1}c_{j,\uparrow,1}$, which generate a quantum of electronic angular momentum while preserving nuclear spin. The mechanism of establishing a $\mathbb{Z}_2$ symmetry is reminiscent of pair hopping of coupled wires~\cite{Kraus2013}, although here it emerges naturally, and thus it is experimentally accessible in a physical setting.

This symmetry is the building tool for the realization of a symmetry-protected quasi-topological phase whose spin sector has the same universal properties of Kitaev's model -- in particular, it hosts MQPs as edge modes. In the following, we discuss the emergence of such phase using a combination of analytical methods and density-matrix-renormalization-group~\cite{White,Schollwoeck} (DMRG) simulations (see Fig.~\ref{fig:scheme}c for typical entanglement spectrum results, as in the Kitaev model). We further elucidate the anyon nature of the edge modes by showing how, upon addition of additional four-body interactions, Eq.~\eqref{H0} can be adiabatically connected to a model with exactly soluble ground state properties~\cite{Iemini2015,Lang2015}, where braiding statistics was recently proved~\cite{Lang2015}.

\paragraph{Low-energy field theory. --} In order to underpin the existence of a quasi-topological phase supporting MQPs as edge modes, we rely on a field theory based on bosonization~\cite{gogolin_book,giamarchi_book}. Within this framework, the essential point is to identify a sector in the low-energy field theory which displays the same physics of Kitaev's chain. Here, we outline the main steps of our treatment (see Ref.~\cite{supmat} for a detailed treatment).
Following conventional bosonization~\cite{gogolin_book,giamarchi_book}, we start by replacing each fermionic mode with a pair of right and left-movers, $c_{j,\alpha, p} = \psi_{\alpha, p; R} (ja) + \psi_{\alpha, p; L} (ja)$, which are given by:
\begin{eqnarray}
\psi_{\alpha, p; r} (x) &=& \frac{\eta_{\alpha, p;r}}{\sqrt{2\pi a}} e^{i \vartheta_{\alpha,p}} \sum_{q}e^{ir qk_{F, \alpha, p}x} e^{-i qr\varphi_{\alpha, p} } 
\end{eqnarray}
with $r=(-1, 1) $ for L/R, and $\varphi_{\alpha, p}$ and $ \vartheta_{\alpha, p}$ being conjugated bosonic operators describing density and phase fluctuations, respectively ($a$ is the lattice spacing). $\eta_{\alpha,p;r}$ are Klein factors, ensuring fermionic commutation relations. 

The low-energy Hamiltonian can then be recast into four (three spin, one charge) sectors. The dynamics in these sectors can be understood after applying two canonical transformations: the first one introduces the bosonic fields $\varphi_{f, S/A} = (\varphi_{\uparrow,f} \pm\varphi_{\downarrow,f})/\sqrt{2}$ (and similarly for $\vartheta_{f, S/A}$), with $f=\pm1$. These bosonic fields describe the behavior of each pair of states coupled by $H_{\textrm{so}}$: in particular, $\vartheta_{1, S/A}$ and $\vartheta_{-1, S/A}$ describe the $\{c_{\uparrow, 1},c_{\downarrow, -1}\}$ and $\{c_{\uparrow, -1},c_{\downarrow, 1}\}$ pair, respectively. The second transformation considers combinations of these fields in the form:
\begin{equation}
\varphi_{f, I} = \frac{\kappa\varphi_{f,S}+\vartheta_{f, A}}{\sqrt{2\kappa}}, \quad\varphi_{f, II} = \frac{\kappa\varphi_{f,S}-\vartheta_{f, A}}{\sqrt{2}\kappa},
\end{equation}
where $\kappa$ denotes the first harmonic commensurate with the spin-orbit term, and is a function of $k_F$ . After this mapping, one is left with a gapless charge sector, described by the fields $\vartheta_{\rho} = \frac{\vartheta_{1, II}+ \vartheta_{-1, II}}{\sqrt{2}}$, two gapped spin sectors which play no major role in the dynamics, and a third spin sector, described by the field $\vartheta_{\sigma} = \frac{\vartheta_{1, II}- \vartheta_{-1, II}}{\sqrt{2}}$, and a Hamiltonian:
\begin{eqnarray}\label{Hsigma}
H_{\sigma}& =&\frac{v_\sigma}{2}\int dx \left[ \frac{(\partial_x\varphi_{\sigma})^2}{K_{\sigma}} + K_{\sigma}(\partial_x\vartheta_{\sigma})^2  \right] +\nonumber\\
&+&\mathcal{W}\int dx \cos[\sqrt{8\pi}\vartheta_{\sigma}] 
\end{eqnarray}
with $\mathcal{W}\propto W$, and $v_\sigma, K_\sigma$ the sound velocity and Luttinger parameter, respectively. This Hamiltonian describes the low-energy physics of the Kitaev model, to which it can be mapped exactly at the Luther-Emery point $K_\sigma =2$~\cite{cheng2011,Kraus2013}. It supports a gapless phase for $K_\sigma\leq 1$, and a gapped, topological phase for $K_\sigma>1$. In the latter, there are two degenerate ground states under open boundary conditions, labeled by different mutual parities $P=\pm1$: this is possible since the model exhibit a $\mathbb{Z}_2$ symmetry, which serves as a symmetry protection mechanism for the quasi-topological phase~\cite{supmat}.

\begin{figure}[t]
\includegraphics[scale=0.223]{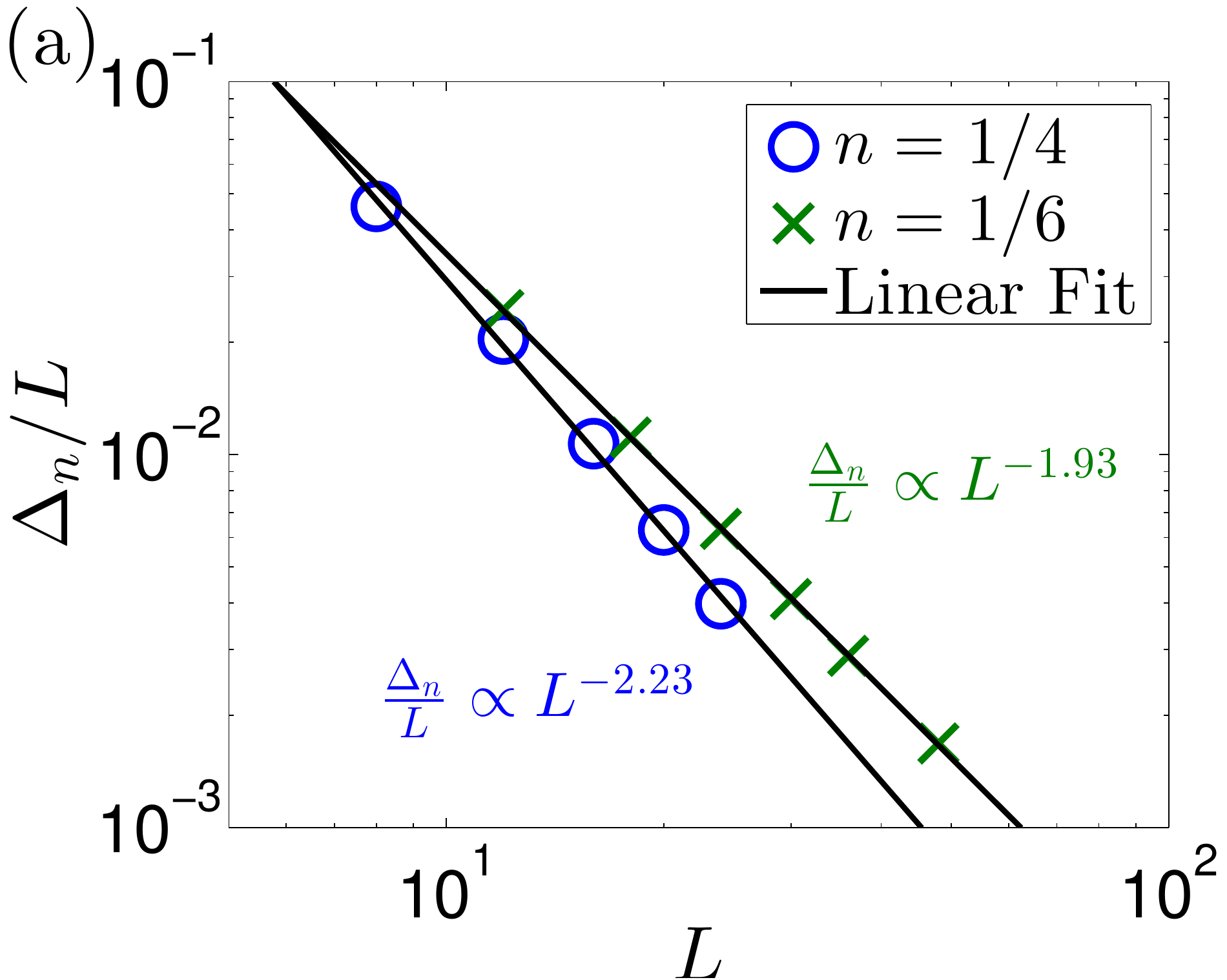}
\includegraphics[scale=0.223]{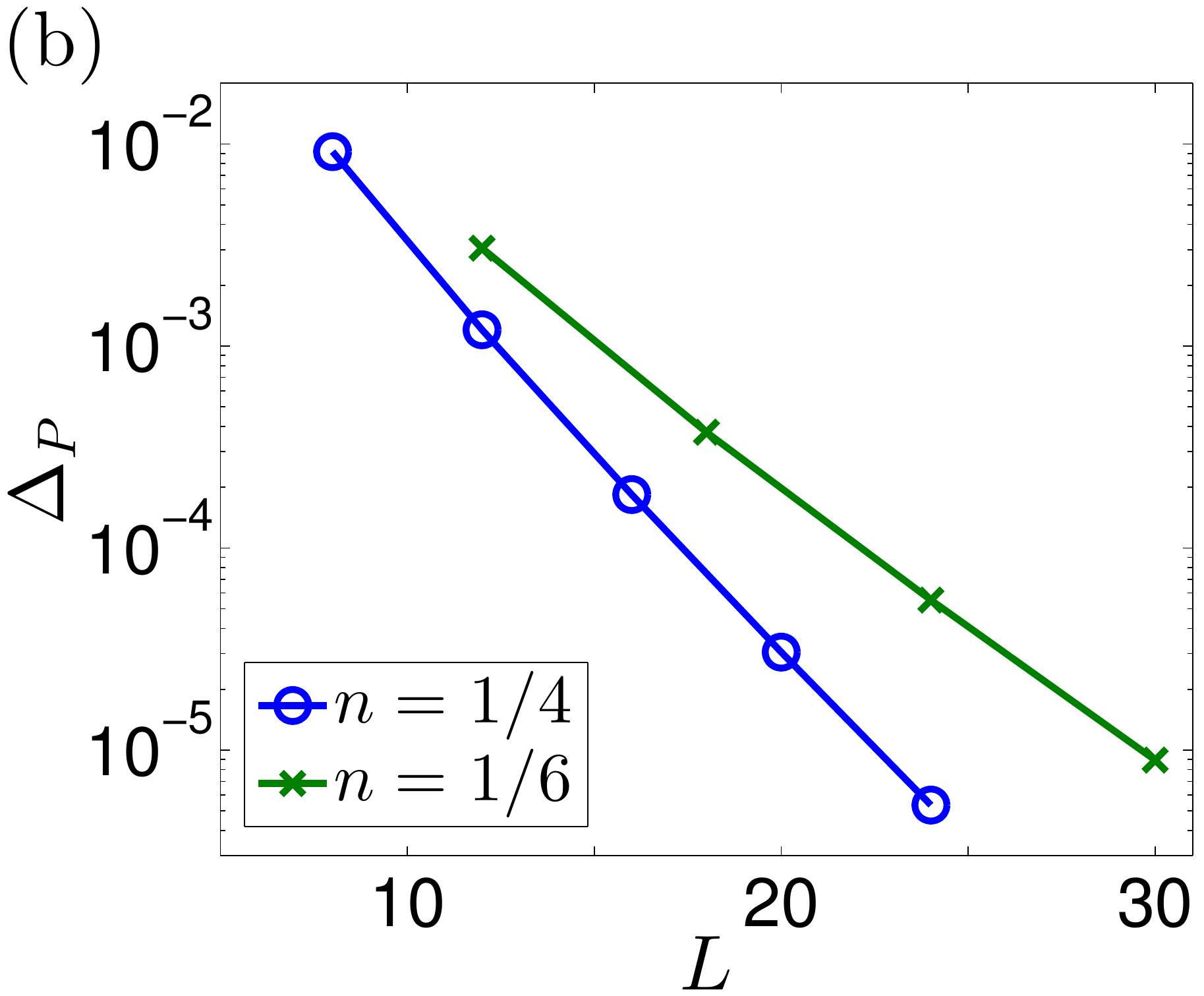}
\includegraphics[scale=0.22]{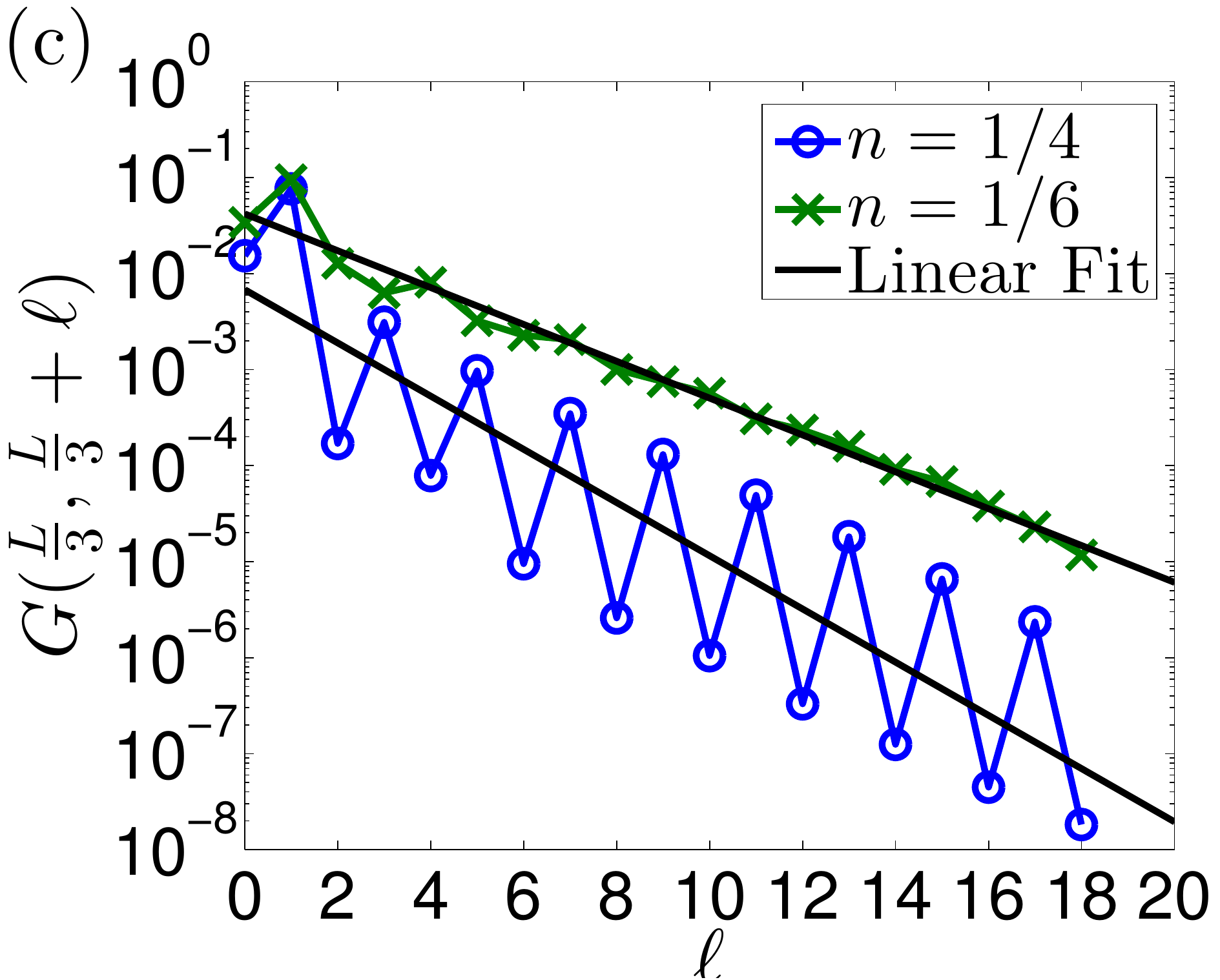}
\includegraphics[scale=0.22]{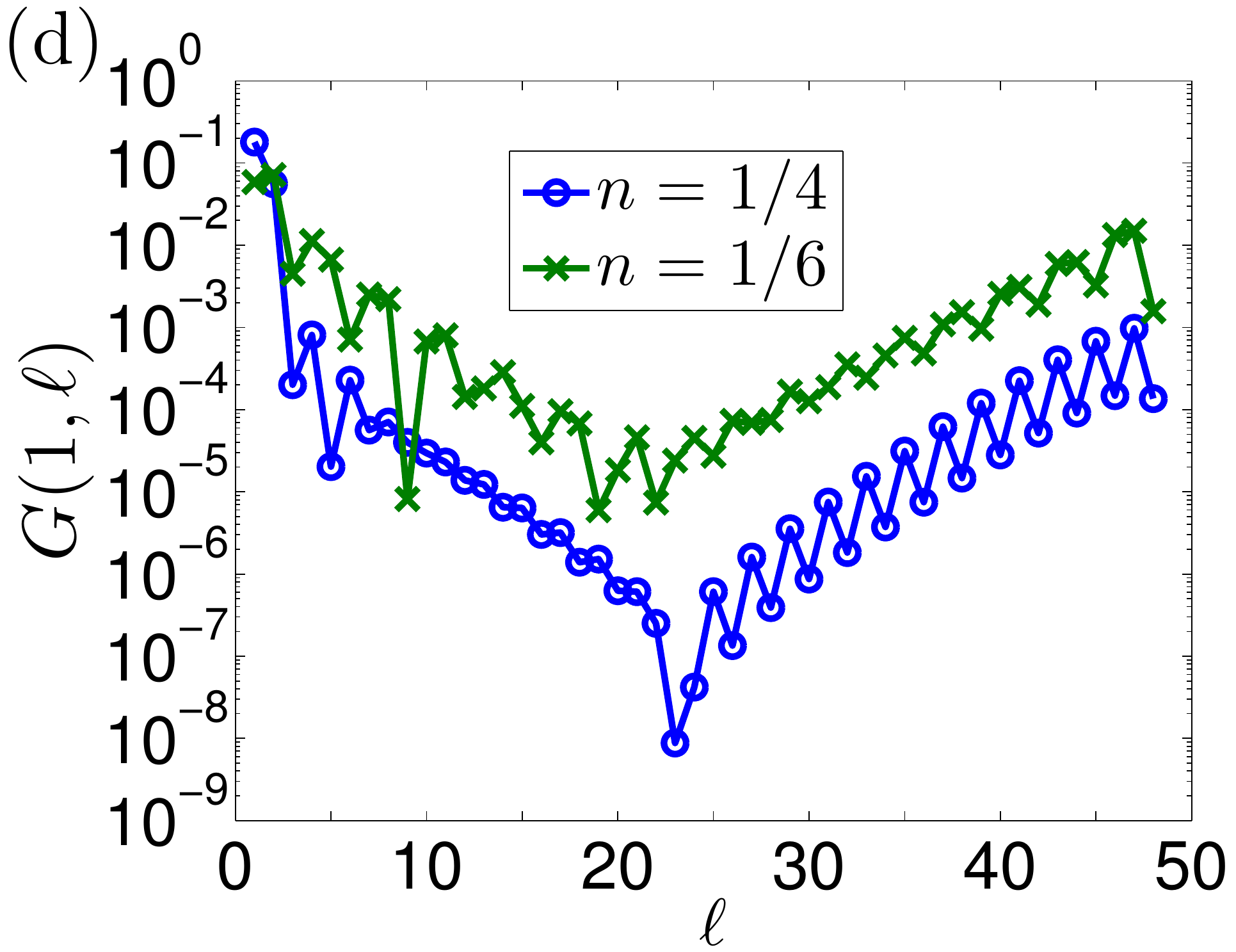}
\centering
\caption{
DMRG analysis of the topological properties of the ground state 
 for 
model $ H$, at fixed parity, 
 with parameters $W=-8,\, \alpha_R=b=4,\, U=0$, at distinct fillings
  $n=1/4$ 
 and $n=1/6$.
   \textbf{(a)} Algebraic scaling of the gap computed at fixed
parity, compatible with $\sim L^{-1}$. 
  \textbf{(b)} Exponential scaling of the gap between the distinct 
 parity sectors.
 \textbf{(c-d)} Single particle correlations $G(j,\ell) = \langle  c^\dagger_{j,\uparrow,1} c_{\ell,\uparrow,1} \rangle$ at the bulk (c) and at the edges (d),
  for a system
  with $L=48$ sites.}
\label{fig:num1}
\end{figure} 

In summary, for sufficiently large $W/t$, the model in Eq.~\eqref{H0} supports a quasi-topological phase, with gapless charge excitations, and decoupled gapped spin-excitations describing a ground state of a Kitaev model, thus supporting MQPs. The role of additional, diagonal interactions can affect the spin sector~\cite{supmat}: since $K_\sigma-1\simeq -(W+U)$, attractive interactions further stabilize the quasi-topological phase, while repulsive interactions require larger values of $W$ to open a gap in the spin sector. Equipped with the guideline provided by the low-energy field theory, we present in the following a non-perturbative analysis of the model based on numerical simulations.

\paragraph{DMRG results. --} In order to demonstrate the existence of a symmetry-protected quasi-topological phase supporting MQPs as edge modes, we employ DMRG simulations based on a rather general decimation prescription for an efficient truncation of the Hilbert space. Typically, we use up to $m=140$ states, which ensure converge on all observables of interest over all parameter regimes~\cite{supmat}. Following the theoretical discussion above, our analysis is based on four observables: {\it (i)} degeneracies in the entanglement spectrum; {\it (ii)} finite-size scaling of energy gaps; {\it (iii)} bulk decay of correlation functions; and {\it (iv)} edge-to-edge correlations. For convenience, we set $t=1$ as energy unit. 

Given the reduced density matrix $\rho_\ell$ with respect of a bipartition of the system cutting the $\ell$-th link of the lattice, the entanglement spectrum is the collection of its eigenvalues $\{\lambda_\alpha\}$, and is known to provide striking signatures of topological order via degeneracies~\cite{Pollmann,Turner}. In Fig.~\ref{fig:scheme}c, we show typical results for the entanglement spectrum in the quasi-topological phase at the representative point $W=-8,\, \alpha_R=b=4,\, U=0$ (these features are stable in a broad parameter range~\cite{MDM2017}). Indeed, the low-lying spectrum displays robust degeneracies for both $n=N/4L=1/4 $ and $n=1/6$ (with $N$ and $L$ total numbers of particles and sites, respectively), as expected for a topological phase supporting MQPs edge modes.

In Fig.~\ref{fig:num1}a, we show the decay of the fixed parity gap with open boundary conditions (OBCs), defined as:
\begin{equation}
\Delta_{n} = E_L^{1}[N, P] - E_L^{0}[N, P]
\end{equation}
where $E_L^n[N,P]$ denotes the $n$-lowest-energy state at size $L$ with number of particles $N$ and mutual parity $P$. The ground state, with energy $E_L^{0}[N, P]$, is always in the $P=1$ sector. In the quasi-topological phase, this gap should decay algebraically due to the presence of a gapless charge excitation. This is confirmed by the DMRG results, as shown in Fig.~\ref{fig:num1}a. Instead, the parity gap:
\begin{equation}
\Delta_{P} = E_L^{0}[N, -1] - E_L^{0}[N, 1]
\end{equation}
is sensitive exclusively to spin excitations. As such, it closes exponentially with the system size $L$, exactly as in the Kitaev chain, as shown in Fig.~\ref{fig:num1}b.

The presence of a finite bulk gap in the spin sector is signalled by an exponential decay of the Green functions, e.g. $G(j, \ell)= \langle c^{\dagger}_{j,\uparrow,1}c_{\ell, \uparrow, 1}\rangle$, in the bulk~\cite{giamarchi_book}.  This is portrait in Fig.~\ref{fig:num1}c, which shows that coherence is rapidly lost as a function of distance in the bulk.

Crucially, the Green functions are also sensitive to the presence of MQPs edge modes, as these operators locally switch parity. In Fig.~\ref{fig:num1}d, we show the correlation of one boundary site with the rest of the chain, $G(1, \ell)$. While the correlation rapidly decays in the bulk due to the presence of a spin gap, there is a strong revival close to the edge of the system, signalling the presence of MQP edge modes. We note that the edge-edge correlation is considerably stronger for filling fractions away from commensurate densities, where the presence of additional (albeit irrelevant) operators is expected to slightly degrade the edge modes, as observed in the Kitaev wire in the presence of repulsive interactions~\cite{stoudenmire2011,tezuka2012}.

\begin{figure}
\includegraphics[scale=0.22]{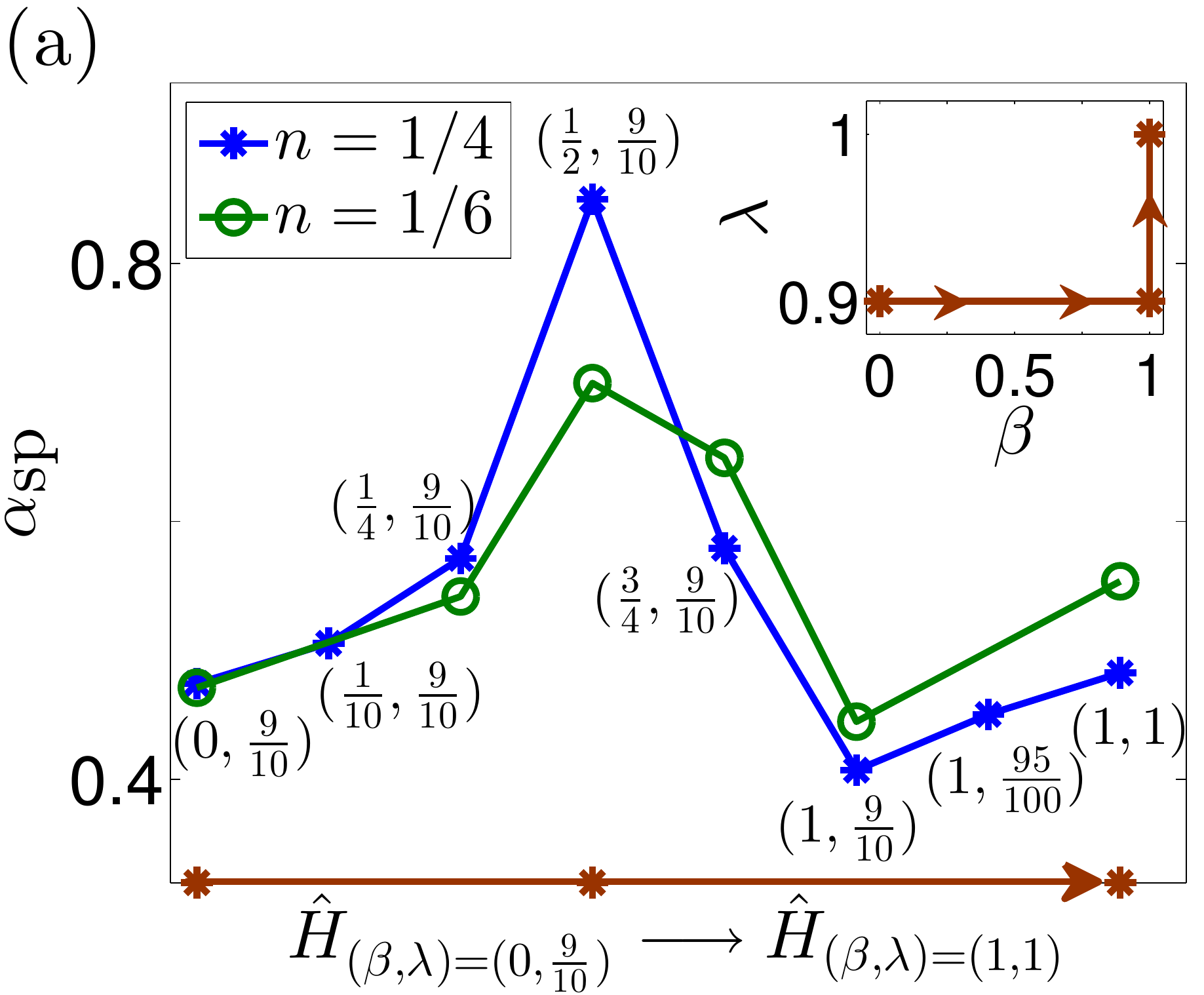}
\includegraphics[scale=0.22]{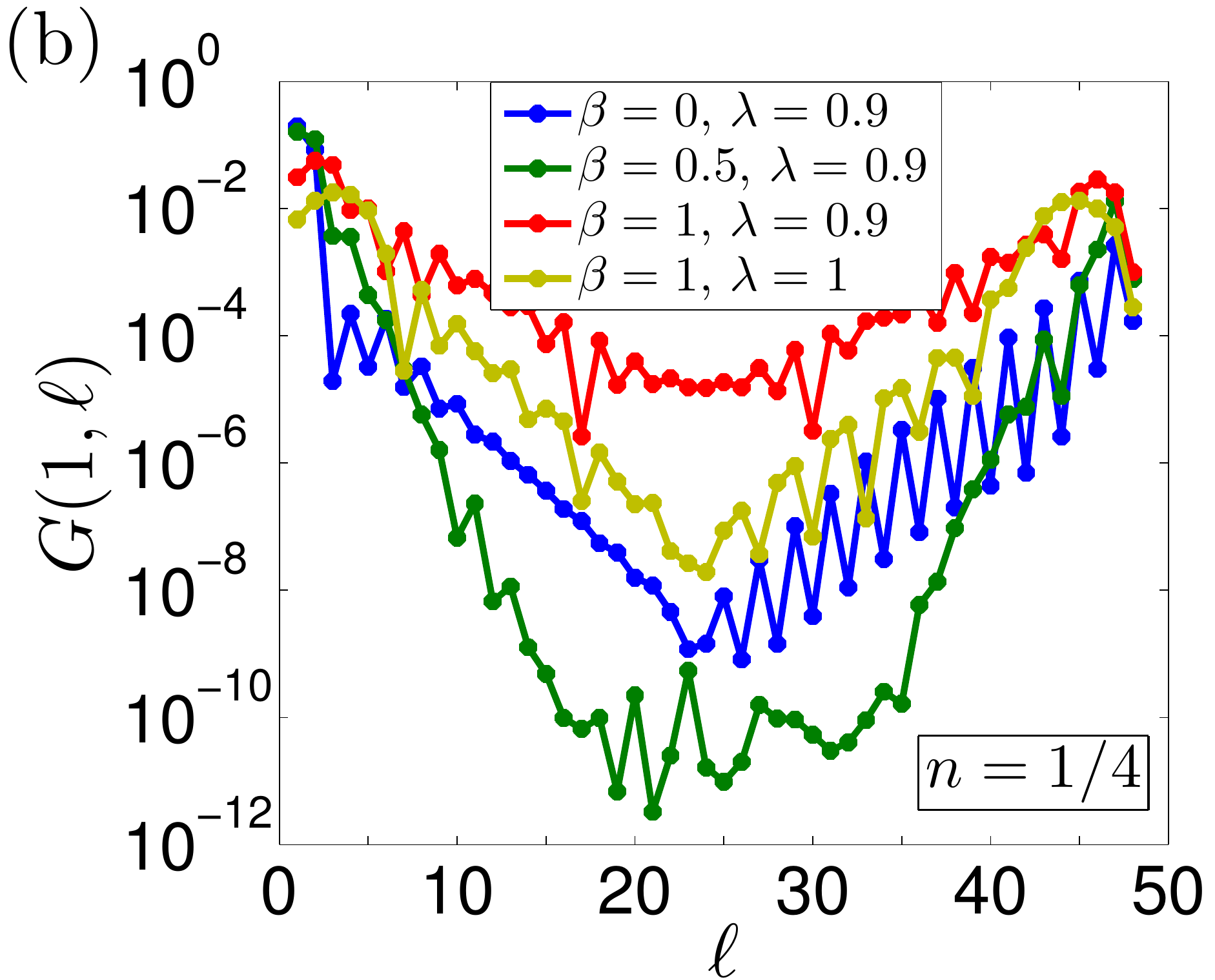}
\centering
\caption{ DMRG analysis for the 
 adiabatic continuity of model $ H$ to an exactly solvable model.
   \textbf{(a)} Parity gap ``$\alpha_{\rm sp}$'' 
  along the adiabatic continuity path $(\beta=0,\, \lambda=9/10) \rightarrow 
  (\beta=1,\, \lambda=1)$. Inset illustrates the Hamiltonian parameters 
 varied along the path.
    \textbf{(b)} Single particle edge correlations 
 along the 
  adiabatic path - similar behavior follows for $n=1/6$.
In all plots we consider a system with $L=48$ sites
and use $m=140$ number of kept states in the DMRG simulations.}
\label{fig:num2}
\end{figure}

\paragraph{Adiabatic continuity of the ground state to an exactly-solvable point. --} Remarkably, it is possible to provide direct evidence for the MQP nature of the edge states by showing how the quasi-topological phase discussed above is adiabatically connected to toy model of spinless fermions with exactly solvable ground state properties~\cite{Iemini2015,Lang2015}, where Ising anyon braiding was recently demonstrated~\cite{Lang2015}. 

The strategy to show adiabatic continuity, discussed in detail in Ref.~\cite{supmat}, consists of three steps. First, for each pair of states coupled by spin-orbit coupling, we restrict the dynamics to the lower band, following a procedure introduced in Ref.~\cite{Budich_2015, FootNote4}. Then, coupling between the lowest bands is introduced via the spin-exchange interaction, and additional four-Fermi couplings. This enlarged Hamiltonian is characterized by two parameters $(\beta,\lambda)$: the point (0, 0.9) represent the model studied in the previous section, while the points $(1,0.9)$ and $(1,1)$ represents two points in the phase diagram of the exactly solvable model~\cite{Iemini2015}. We note that all symmetries of the problem are kept for arbitrary $(\beta,\lambda)$.

Within this enlarged parameter space, we have carried out DMRG simulations to show that the gap in the spin sector does not close. The latter was extracted from the decay of the Green function in the bulk, $G(j,\ell)\simeq e^{-\alpha_{sp}|j-\ell |}$, and is depicted in Fig.~\ref{fig:num2}a. Along the full path in parameter space, the gap stays open, implying that out quasi-topological state is the same phase as in Ref.~\cite{Iemini2015}. Another striking signature of adiabatic continuity is the fact that all diagnostics applied before signal topological order all along the path. This is illustrated in Fig.~\ref{fig:num2}b, where we plot the edge-edge Green function at several points along the path itself.

\paragraph{Realization using alkaline-earth-like atoms in optical lattices. -} The model discussed above finds a natural implementation using fermionic isotopes of AEA in optical lattices~\cite{cazalilla2014}, such as $^{171}$Yb, $^{173}$Yb, and $^{87}$Sr. As illustrated in Fig.~\ref{fig:scheme}, the orbital degree of freedom is encoded in the electronic state: the $^1$S$_0$ ground state manifold representing $p=-1$, and the long-lived excited state manifold $^3$P$_0$ representing $p=1$. The spin degree of freedom is instead encoded in the nuclear spin state, which for AEA is basically decoupled from the electronic degree of freedom for both ground and low-lying excited states. In $^{171}$Yb, the nuclear spin is $I=1/2$, so all degrees of freedom are immediately available as required here. For $^{173}$Yb and $^{87}$Sr, which do have $I=5/2$ and $9/2$, respectively, unwanted Zeeman states can be excluded from the dynamics either employing state-dependent light-shifts~\cite{mancini2015}, or by exploiting the fact that the clock frequency is $m_F$-dependent due to different linear Zeeman shifts in the $^1$S$_0$ and $^3$P$_0$ manifold. 

The two key elements of our proposal, large spin-exchange interactions and spin-orbit couplings on the so-called clock transition, build upon state-of-the-art experimental progresses in AEA physics. As demonstrated in recent experiments with $^{173}$Yb~\cite{cappellini2014,scazza2014}, the spin-exchange interaction in these settings can be extremely large, of the order of $5/10$ kHz in optical lattices, guaranteeing that the driven interaction strength in our system is considerably larger than typical temperatures. Moreover, the ratio $W/t$ can be tuned via modifying either the optical lattice depth or the trapping in the transverse direction. Spin-orbit coupling between ground and excited states has been recently demonstrated both at JILA~\cite{kolkowitz2016} and at LENS~\cite{livi2016} realizing single-particle band structures akin to the one employed here, albeit with slightly different microscopic Hamiltonians (following Ref.~\cite{Budich_2015}, the precise form we use here requires a tilting of the lattice or a superlattice structure). In concrete, considering typical tunneling rates of order $t\simeq 100h$ Hz, spin-orbit couplings of order $400h$ Hz and spin-exchange interactions of order $800h$ Hz would give direct access to the quasi-topological phase. 
In these experimental settings, the quasi-topological phase can be characterized using both correlation function and spectral properties, as discussed above. The nature of the edge modes can be demonstrated using a variety of techniques~\cite{kraus2012}. In particular, time-of-flight imaging and edge spectroscopy can be used to demonstrate the existence of zero-energy modes and their inherent correlations. Moreover, the fact that our model is adiabatically connected to an exactly-solvable point provides a qualitative guidance on the shape of the MQP wave-function - generically hard to analytically access in interacting systems -, opening up a concrete perspective to realize braiding operations in such settings.

\paragraph{Conclusions. -} We have shown how Majorana quasi-particles can emerge as edge modes of orbital Hubbard models in the presence of spin-orbit interactions. The key element for the realization of the quasi-topological phase supporting them is angular momentum conservation, an epitome building block of atomic physics experiments. The stability of the mechanism we propose paves the way toward the investigation of interacting topological states and Majorana edge modes in both atomic clocks and optical lattice experiments, where the main ingredients of our proposal are naturally realized and have been experimentally demonstrated over the last two years.

\paragraph{Acknowledgement. - }

We acknowledge useful discussions with M. A. Baranov, J. Catani, D. Rossini, and C. Sias. This work is partly supported by EU-IP-QUIC (R.~F.), the ERC Synergy grant UQUAM (P.~Z.), and the ERC Consolidator grant TOPSIM (L.~F.). 
L.~M. was supported by LabEX ENS-ICFP: ANR-10-
LABX-0010/ANR-10-IDEX-0001-02 PSL*. 
The numerical part of this work has been performed using the DMRG code  released  within  the `Powder with Power` project (http://qti.sns.it/dmrg/home.html). This work was granted access to the HPC resources of MesoPSL financed
by the Region Ile de France and the project Equip@Meso (reference ANR-10-EQPX-29-01).

{\it Note added.} While completing this work, a pre-print appeared~\cite{zhou2016}, where the commensurate regime of a model combining spin-exchange interactions with a different type of spin-orbit coupling was investigated.

\clearpage
\newpage

\clearpage
\setcounter{equation}{0}%
\setcounter{figure}{0}%
\setcounter{table}{0}%
\renewcommand{\thetable}{S\arabic{table}}
\renewcommand{\theequation}{S\arabic{equation}}
\renewcommand{\thefigure}{S\arabic{figure}}

\onecolumngrid

\begin{center}
  {\Large Supplemental Material for: \\ 
  Majorana Quasi-Particles Protected by $\mathbb{Z}_2$ Angular Momentum Conservation}

\end{center}

In this Supplemental Material we provide additional information 
on the low-energy field theory, on the adiabatic continuity to a model with exactly solvable ground state properties, and on the numerical simulations.

\section{Low-energy field theory} 

Here, we discuss a bosonization approach to address the low-energy physics of $H$. In principle, it is possible to employ as a starting point either $H_t$, or $H_{t}+H_{SO}$: in both cases, the topological phase discussed in the main text is found. In the following, we follow the former approach, which has the advantage of providing a simpler description of the effects of the diagonal interaction on the stability of the phase.

We first replace the fermionic operators with right and left-movers:
\begin{equation}
c_{j,\alpha, p} = \psi_{\alpha, p; R} (x = ja) + \psi_{\alpha, p; L} (x = ja)
\end{equation}
where $a$ is the lattice spacing, and then introduce the conventional bosonization representation:
\begin{eqnarray}
\psi_{\alpha, p; r} (x) &=& \frac{\eta_{\alpha, p;r}}{\sqrt{2\pi a}} e^{ir k_{F, \alpha, p}x} e^{-i (r\varphi_{\alpha, p} -\vartheta_{\alpha,p})} 
\end{eqnarray}
with $r=(-1, 1) $ for L/R, and $(\varphi_{\alpha, p}, \vartheta_{\alpha, p})$ being conjugated bosonic operators describing density and phase fluctuations, respectively, and $\eta_{\alpha, p;r} $ are Klein factors (neglected in the following, they play a similar role as in Ref.~\cite{SM_Kraus2013}). For our purposes, it would be useful to consider the generic form of the operators including all harmonics, i.e.:
\begin{eqnarray}
\psi_{\alpha, p; r} (x) &=& \frac{\eta_{\alpha, p;r}}{\sqrt{2\pi a}} e^{i \vartheta_{\alpha,p}} \sum_{q}e^{ir qk_{F, \alpha, p}x} e^{-i qr\varphi_{\alpha, p} } 
\end{eqnarray}
We take as our starting point $H_{t}$. For convenience, we introduce two independent sector $f = -1, 1$, with two pair of symmetric and antisymmetric fields, $\varphi_{f, S/A} = (\varphi_{\uparrow,f} \pm\varphi_{\downarrow,f})/\sqrt{2}$.
In each of these two sectors, the low-energy physics is described by the conventional theory of spin-orbit coupled gases, given by
\begin{eqnarray}\label{Hf.SM}
H_{f}& =& \sum_{Z=S, A}\frac{v_Z}{2}\int dx \left[ \frac{(\partial_x\varphi_{f,Z})^2}{K_{f,Z}} + K_{f,Z}(\partial_x\vartheta_{f,Z})^2  \right] +\nonumber\\
&+&g_{SO}\int dx \cos[\sqrt{2\pi}(\vartheta_{f,A} +\kappa\varphi_{f,S})] 
\end{eqnarray}
where the Luttinger parameters are all set to 1, and $\kappa$ denotes the first harmonic commensurate with spin orbit coupling, which is a function of $k_F$.

After defining new fields, 
\begin{equation}
\varphi_{f, I} = (\kappa\varphi_{f,S}+\vartheta_{f, A})/\sqrt{2\kappa}, \quad\vartheta_{f, I} = (\vartheta_{f,S}+\kappa\varphi_{f, A})/\sqrt{2\kappa}
\end{equation}
and 
\begin{equation}
\varphi_{f, II} = (\kappa\varphi_{f,S}-\vartheta_{f, A})/\sqrt{2}\kappa, \quad\vartheta_{f, II} = (\vartheta_{f,S}-\kappa\varphi_{f, A})/\sqrt{2}
\end{equation}
one can infer from Eq.~\ref{Hf.SM} that the $\varphi_{f,I}$ are gapped by the cosine terms, while the $\varphi_{f, II}$ remain gapless. In this regime, we can define collective charge and spin fields:
\begin{equation}
\vartheta_{\rho} = \frac{\vartheta_{1, II}+ \vartheta_{-1, II}}{\sqrt{2}}, \qquad \vartheta_{\sigma} = \frac{\vartheta_{1, II}- \vartheta_{-1, II}}{\sqrt{2}}
\end{equation}
whose physics is, in the absence of interactions, described by two decoupled Tomonaga-Luttinger liquids (TLLs). In this basis, the spin-exchange term reads:
\begin{eqnarray}
H_{W,j} &\simeq & W\int dx \left[e^{i\sqrt{2\pi}\vartheta_{1, S}}\sum_{q}\sin(\sqrt{2\pi}q\varphi_{1, A})  \right] \times        
\left[e^{-i\sqrt{2\pi}\vartheta_{-1, A}}\sum_{q}\sin(\sqrt{2\pi}q\varphi_{-1, S}) \right]+\text{h.c.}  \nonumber\\ 
& \equiv & W\int dx  e^{i\sqrt{2\pi}(\vartheta_{1, S}-\kappa\varphi_{1, A} )}e^{-i\sqrt{2\pi}(\vartheta_{-1, A}-\kappa\varphi_{-1, S} )} + \text{h.c.}+....
\nonumber\\ 
& =& W\int dx  e^{i\sqrt{4\pi}\vartheta_{1, II}}e^{-i\sqrt{4\pi}\vartheta_{-1, I}} + \text{h.c.}+....
\nonumber\\ 
& \simeq &W\int dx\cos\left[ \sqrt{8\pi}\vartheta_{\sigma}   \right].
\end{eqnarray}
where at each stage, the dots indicate the same contributions which are either oscillating, or contain fields which are gapped away from the dynamics, and thus can be neglected. The free Hamiltonian reads:
\begin{eqnarray}
H_{\text{free}}& =& \sum_{f}\frac{v_{II}}{2}\int dx \left[ \frac{(\partial_x\varphi_{f,II})^2}{K_{f,II}} + K_{f,II}(\partial_x\vartheta_{f,II})^2  \right] =\nonumber\\
&=& \frac{v_\rho}{2}\int dx \left[ \frac{(\partial_x\varphi_\rho)^2}{K_{\rho}} + K_{\rho}(\partial_x\vartheta_{\rho})^2  \right] + \frac{v_\sigma}{2}\int dx \left[ \frac{(\partial_x\varphi_\sigma)^2}{K_{\sigma}} + K_{\sigma}(\partial_x\vartheta_{\sigma})^2  \right]
\end{eqnarray}
with:
\begin{equation}
v_{II}K_{II} = \frac{v_S+v_A/\kappa^2}{2},\quad v_{II}/K_{II} = \frac{v_S+v_A\kappa^2}{2}, 
\end{equation}
which implies:
\begin{equation}
v_{\rho}K_{\rho} = \frac{v_S+v_A/\kappa^2}{2},\quad v_{\rho}/K_{\rho} = \frac{v_S+v_A\kappa^2}{2}, 
\end{equation}
\begin{equation}
v_{\sigma}K_{\sigma} = \frac{v_S+v_A/\kappa^2}{2},\quad v_{\sigma}/K_{\sigma} = \frac{v_S+v_A\kappa^2}{2}, 
\end{equation}
For the spin sector, the spin Luttinger parameter is:
\begin{equation}
K_\sigma^{\text{free}} = \sqrt{\frac{1+v_A/v_S}{1+v_A\kappa^4/v_S}}
\end{equation}
which implies that, since $\kappa>1$, the Luttinger parameter gets small for equal velocities. We note here that the scaling dimension of the spin-exchange operator is:
\begin{equation}
d_W = 2 /K_\sigma
\end{equation}
We remark here that both $H_W$ and the diagonal interactions can be used to drastically enhance $K_\sigma$. In particular, one has:
\begin{equation}
K_\sigma = K^{\text{free}}_\sigma - a_1W -a_2 U
\end{equation}
with $a_1, a_2$ being non-universal, positive prefactors which depend on $\kappa,\alpha_R, b$.

In summary, the charge sector remains gapless (away from commensurability points which can introduce additional umklapp terms), while the spin sector is described by the same low-energy field theory of the Kitaev chain. In the bosonized language, the relevance of the $\mathbb{Z}_2$ symmetry is clear: in the absence of it, terms of the form $\cos(\sqrt{4\pi}\vartheta_{\sigma})$ would appear at low energies, and immediately spoil the link to the Kitaev chain.
Furthermore, at the Luther-Emery point $K_\sigma=2$, it is possible to re-fermionize the model so that the correspondence is even clearer - see, e.g, Ref.~\cite{SM_cheng2011,SM_Kraus2013}

  \section{Spectral properties}

\subsection{Full quantum number labelling}

The theory is defined by a pair of quantum numbers:
\begin{equation}
N  = N_1+N_2, \qquad P = (-1)^{(N_1-N_2)/2} = (-1)^{N_1 - N/2}.
\end{equation}
From the bosonization analysis described above, one expects the following scaling in the topological phase for different boundary conditions (periodic (PBC) and open (OBC)):
\begin{equation}
\text{OBC/PBC:}\quad E_\rho = E(N+2, P) + E(N-2, P) - 2E(N, P) \simeq 1/L
\end{equation}
\begin{equation}
\text{OBC:}\quad E_\sigma = E(N, P) - E(N, -P)  \simeq e^{-\alpha L}
\end{equation}  
\begin{equation}
\text{PBC:}\quad E_\sigma = E(N, P) - E(N, -P)  \simeq const.
\end{equation}  
where $E(N,P)$ are the ground state energies at size $L$ in the sector with $N$ number of particles and $P$ mutual parity. In the main text, we have employed the first two scalings above as characteristic signatures of the topological phase in our DMRG simulations.

\section{Adiabatic Continuity to an Exactly Solvable Model}

Remarkably, it is possible to provide further evidence for 
the MQP nature of the edge states by showing how the quasi-topological 
phase discussed in the main text is adiabatically connected to an exactly solvable 
model where Ising anyon braiding was recently demonstrated~\cite{SM_Iemini2015,SM_Lang2015}. 
The adiabatic continuity can be studied in the lower band projected Hamiltonian
in the limit of strong spin-orbit coupling. Specifically, we employ below the generalized spin-orbit coupling introduced in Ref.~\cite{SM_Budich_2013,SM_Budich_2015}, which allows a straightforward, yet exact projection into the lower band of each pair of coupled states. 
In this regime, the non-interacting part
 of the Hamiltonian,
\begin{eqnarray}
{H}_{\rm{non}} &=& \sum_j {H}_{t,j}  + {H}_{so,j},\nonumber \\
&=&   \sum_p \left[ \sum_{j,\alpha} t(c^\dagger_{j, \alpha, p}c_{j+1, \alpha, p}+\text{h.c.}) + 
\sum_j \left\{(\alpha_R+b) \cd{j,\uparrow,p}c_{j+1,\downarrow,-p}
+ (b-\alpha_R) \cd{j+1,\uparrow,p}c_{j,\downarrow,-p}  + \text{h.c.}\right\}  \right]\nonumber \\
&=& \sum_p  h_{\rm{non},p}
\end{eqnarray}
describes two well separated Bloch bands by a Fourier transformation.
 Let us focus for simplicity in only one of the two decoupled parts 
of the non-interacting Hamiltonian, \textit{i.e.}, a single 
term ``$ h_{\rm{non},p}$''.
On Fourier transform we easily obtain its Bloch Hamiltonian,
\begin{eqnarray}
{h}_{\rm{non},p}(k) &=& d^\mu (k)  \sigma_\mu,  \qquad \mu=0,x,y,z \nonumber \\
d^\mu(k) &=& 2\,(t\cos(k), b\cos(k),\alpha_R \sin(k),0)
\end{eqnarray}
with $ \sigma_\mu$ are the usual Pauli matrices. The band structure 
and Bloch functions are explicitly given by,
\begin{eqnarray}
E_{\pm}(k) &=& d^0 \pm |\vec{d}|,\quad \vec{d} = (d^x,d^y,d^z) \nonumber \\
|u_{\pm}(k) \rangle  &=& \frac{{P}_{\pm}(k) |\uparrow \rangle}{|{P}_{\pm}(k) |\uparrow\rangle|},
 \quad  \sigma^z |\uparrow\rangle = |\uparrow\rangle \nonumber \\
 {P}_{\pm}(k) &=& \frac{1}{2} \left( 1 \pm {d}(k) \cdot   \sigma \right), \quad 
 {d} = \frac{\vec{d}}{|\vec{d}|}
\end{eqnarray}

 Considering that the ordinary 
hopping term does not influence the Bloch states due to its spin
 independence, the effective Hamiltonian in the lower band,
 at $\alpha_R=b$, is described by a flat band,
\begin{eqnarray}
{h}_{\rm non,p} &=& 2b \sum_{j} \left[   \gamma_{j,p,+}^\dagger  \gamma_{j,p,+} 
-  \gamma_{j,p,-}^{\dagger}  \gamma_{j,p,-}
 + H.c. \right]  \nonumber \\
  &\sim& -2b \sum_{j} \left[  \gamma_{j,p,-}^\dagger  \gamma_{j,p,-}
    + H.c.\right]
\end{eqnarray}  
     with a gap $4b$ from the upper band, and 
 Bloch states $ \gamma_{j,p,\pm} = ({c}_{j,\downarrow,-p} \pm {c}_{j+1,\uparrow,p})/\sqrt{2}$;
 inversely, 
\begin{equation}
{c}_{j,\downarrow,- p} = \frac{1}{\sqrt{2}}( \gamma_{j,p,+} +  \gamma_{j,p,-}), \qquad
{c}_{j,\uparrow, p} = \frac{1}{\sqrt{2}}( \gamma_{j-1,p,+} -  \gamma_{j-1,p,-})
\end{equation} 

We notice now that interacting terms, such as spin-exchange interactions, are
  effectively described in the lower band picture as pairing interactions 
  (for simplicity of notation we use hereafter the operators
  $ \eta_{j} \equiv  \gamma_{j,1,-}$ and $ \chi_{j} \equiv  \gamma_{j,-1,-}$
   to explicitly represent the lower band subspace), 
\begin{eqnarray} \label{eq:sup:lwband:HW}
 H_{W,j} &=& W_{\text{ex}} (c^\dagger_{j, \uparrow, -1}c^\dagger_{j, \downarrow, 1} c_{j, \downarrow, -1} c_{j, \uparrow, 1} + \text{h.c.}) \nonumber \\
&=& (W_{\rm ex}/4)\, ( \gamma_{j-1,-1,+}^\dagger -  \gamma_{j-1,-1,-}^\dagger) 
( \gamma_{j,-1,+}^\dagger +  \gamma_{j,-1,-}^\dagger )
( \gamma_{j,1,+} +  \gamma_{j,1,-}) 
( \gamma_{j-1,1,+} -  \gamma_{j-1,1,-})  \nonumber \\
    &\sim &  (W_{\rm ex}/4)\,( \chi_{j}^\dagger \chi_{j-1}^\dagger   \eta_{j-1}  \eta_{j} +
   \text{h.c.} )
\end{eqnarray} 
a key element in order to generate MQPs according to previously 
studied models~\cite{SM_Kraus2013,SM_Iemini2015,SM_Lang2015}. 

 Within such a lower band picture one may also introduce some additional
  Hamiltonian terms ${H}_{\rm{loc}U,j} = U_{\rm{loc}} \sum_p ( n_{j,\uparrow,p} + 
   n_{j,\downarrow,-p})^2$, and 
  ${H}_{\rm{lcso},j} = h_{\rm{lcso}} (\cd{j,\uparrow,1}c_{j,\downarrow,-1} +
   H.c.)(\cd{j,\uparrow,-1}c_{j,\downarrow,1} + H.c.)$
 describing 
 local diagonal interactions and a local coherent spin-orbit coupling, respectively,
   in such a way to 
 reproduce the exactly solvable model $ {H}_{\lambda}$ 
  proposed in \cite{SM_Iemini2015}. 
  Briefly recalling, the model in Ref.~\cite{SM_Iemini2015} is given by,
  \begin{eqnarray}
  \label{eq:exact.solvable.hamiltonian}
\frac{ H_\lambda}{4} &=& - \sum_{\substack{j=1 , \alpha =\chi,\eta}}^{L-1} \Big[ ( \alpha^\dagger_{j} \alpha_{j+1} \! + \! \text{H.c.}) - \! ( n_j^{\alpha} +  n_{j+1}^{\alpha}) + \! \lambda  n_j^{\alpha}  n_{j+1}^{\alpha} \Big] + \,\nonumber \\
&  & - \frac{\lambda}{2} \sum_{j=1}^{L-1} \Big[ ( n_j^\chi +  n_{j+1}^\chi)( n_j^\eta +  n_{j+1}^\eta) - ( \chi^\dagger_{j} \chi_{j+1} +  \chi^\dagger_{j+1} \chi_{j} ) ( \eta^\dagger_{j}  \eta_{j+1} 
+ \eta^\dagger_{j+1}  \eta_{j}) + 2 \,( \eta^\dagger_{j}  \eta^\dagger_{j+1}  \chi_{j+1}  \chi_{j} + {\rm H.c.}) \Big] 
\end{eqnarray} 
where 
$ n_j^\chi =  \chi_j^\dagger  \chi_j$, 
$ n_j^\eta =  \eta_j^\dagger  \eta_j$.   
  The model has an exactly solvable line at $\lambda=1$,
   on varying the density of particles, described by a topological 
   nontrivial ground state. Away from the exactly solvable line the system is 
   topological for $\lambda < 1$, while describing a phase separation (PS) state 
   for $\lambda>1$.
   
We will now explicitly show that the additional Hamiltonian terms ${H}_{\rm{loc}U,j}$, 
${H}_{\rm{lcso},j}$
 with the diagonal and spin-exchange interactions, ${H}_U$ and ${H}_{W_{\rm ex}}$,
  as described in the main text, 
 indeed reproduce the exact solvable model in their lower band picture. 
 
 The diagonal
 interactions 
 ${H}_{U,j} = \sum_{p} U_{p}  n_{j,\uparrow,p}  n_{j,\downarrow,p} +  U\sum_{\alpha, \beta}  n_{j,\alpha,-1}  n_{j,\beta,1}$,
  are described 
  in the lower band at $U_p=U$ as follows,
  \begin{eqnarray} \label{eq:sup:lwband:HU}
{H}_{U,j} &=& (U/2) \sum_{\alpha,p}\left\{  \left( n_{j,\alpha,p} \right)^2 - \left( n_{j,\alpha,p} \right) \right\} \nonumber \\
&\sim &  \frac{U}{4}\left[
\left(  n_j^{\chi} n_{j-1}^{\chi} +  n_j^{\eta} n_{j-1}^{\eta}   \right)
 +  \left(  n_j^{\chi} +  n_{j-1}^{\chi}\right) 
 \left(  n_j^{\eta} +  n_{j-1}^{\eta} \right) \right]
  \end{eqnarray}
 
 The two additional Hamiltonian terms are described in the lower band as,
 \begin{eqnarray} \label{eq:sup:lwband:HlocU:Hcso}
 H_{\rm locU,j} &=& \frac{U_{\rm loc}}{4}  ( n_j^\chi +  n_{j-1}^\chi + 
  n_j^\eta +  n_{j-1}^\eta) + \frac{U_{\rm loc}}{2} ( n_j^\chi  n_{j-1}^\chi +
   n_j^\eta  n_{j-1}^\eta) \\
  H_{\rm cso,j} & \sim & \frac{h_{\rm cso}}{4} \, ( \chi_j^\dagger \chi_{j-1} + H.c.)
( \eta_j^\dagger \eta_{j-1} + H.c.)
 \end{eqnarray}

 Thus, one can see by Eqs.\eqref{eq:sup:lwband:HW},\eqref{eq:sup:lwband:HU},\eqref{eq:sup:lwband:HlocU:Hcso} that the total Hamiltonian,
 $ H = \sum_j ({H}_{\rm{so},j} + {H}_{t,j}  + {H}_{U,j} + {H}_{W,j} +  {H}_{\rm{lcso},j} + {H}_{\rm{loc}U,j})$, in the lower band picture ($\alpha_R=b \gg$ other terms), 
 describes the exactly solvable model (Eq.\eqref{eq:exact.solvable.hamiltonian}) with:
 $W_{\rm ex}/t=-4\lambda$, $U_p=U=2U_{\rm loc}=-2\lambda t$, and $h_{\rm clso}/t = 2\lambda$.

\paragraph{DMRG results on the adiabatic continuity. - } We perform our DRMG analysis for the adiabatic continuity in two steps:
  i) first connect our initial \textit{physical} Hamiltonian with
  $\alpha_R=b=8t$, $W_{\rm ex}/t=-8 $, $U_p=U=U_{\rm loc}=h_{\rm clso} = 0$ to the
  model $\hat{H}_{\lambda}$ at $\lambda=0.9$; ii)
  connect $\hat{H}_{\lambda}$ from $\lambda=0.9$ to the exactly solvable line $\lambda=1$~.
  We choose to perform the adiabatic continuity in such two steps
  in order to avoid possible phase transitions along the path to the
  exactly solvable line $\lambda=1$.
  The total Hamiltonian is thus parametrized by $\hat{H}_{(\beta,\lambda)}$ as follows:
\begin{eqnarray}
W_{\rm{ex}}/t &=& (-8 + 4.4\beta)\,(\lambda/0.9) \nonumber \\
U_p/t &=& U/t = 2 U_{\rm{loc}}/t = -2\beta \lambda \\
h_{\rm clso}/t &=& 2\beta \lambda \nonumber
\end{eqnarray}
where $(\beta=0,\, \lambda=0.9) \rightarrow (1,0.9)$ describes the first step, and
$(1,0.9) \rightarrow (1,1)$ the second one.
 Our results are presented in Fig.~3a of the main text, where we show
 that
the parity gap ``$\alpha_{\rm sp}$'',
 computed from the exponential scaling of
    the Green functions in the bulk ($G(j,\ell) \sim e^{-\alpha_{\rm sp} |j-\ell|}$),
 does not close along the adiabatic path.
In Fig.~3b we see the persistence of edge-edge correlations
along the path. It was also verified that the entanglement spectrum degeneracy is kept
  for all parameters. Finally, as expected, for $\lambda>1$ all of these topological
  properties are absent, and one observes a PS state (not shown).

\section{Convergence of the Numerical Simulations}

In this section we briefly present the analysis for the numerical accuracy of our results.
We employ state-of-the-art DMRG simulations based on a rather general decimation prescription for an efficient truncation of the Hilbert space. Typically, we observed that using
 $m \sim 180$ states in the DRMG simulations, already ensure negligible truncation errors 
 for the observables under analysis.
 
\begin{figure}
\includegraphics[scale=.5]{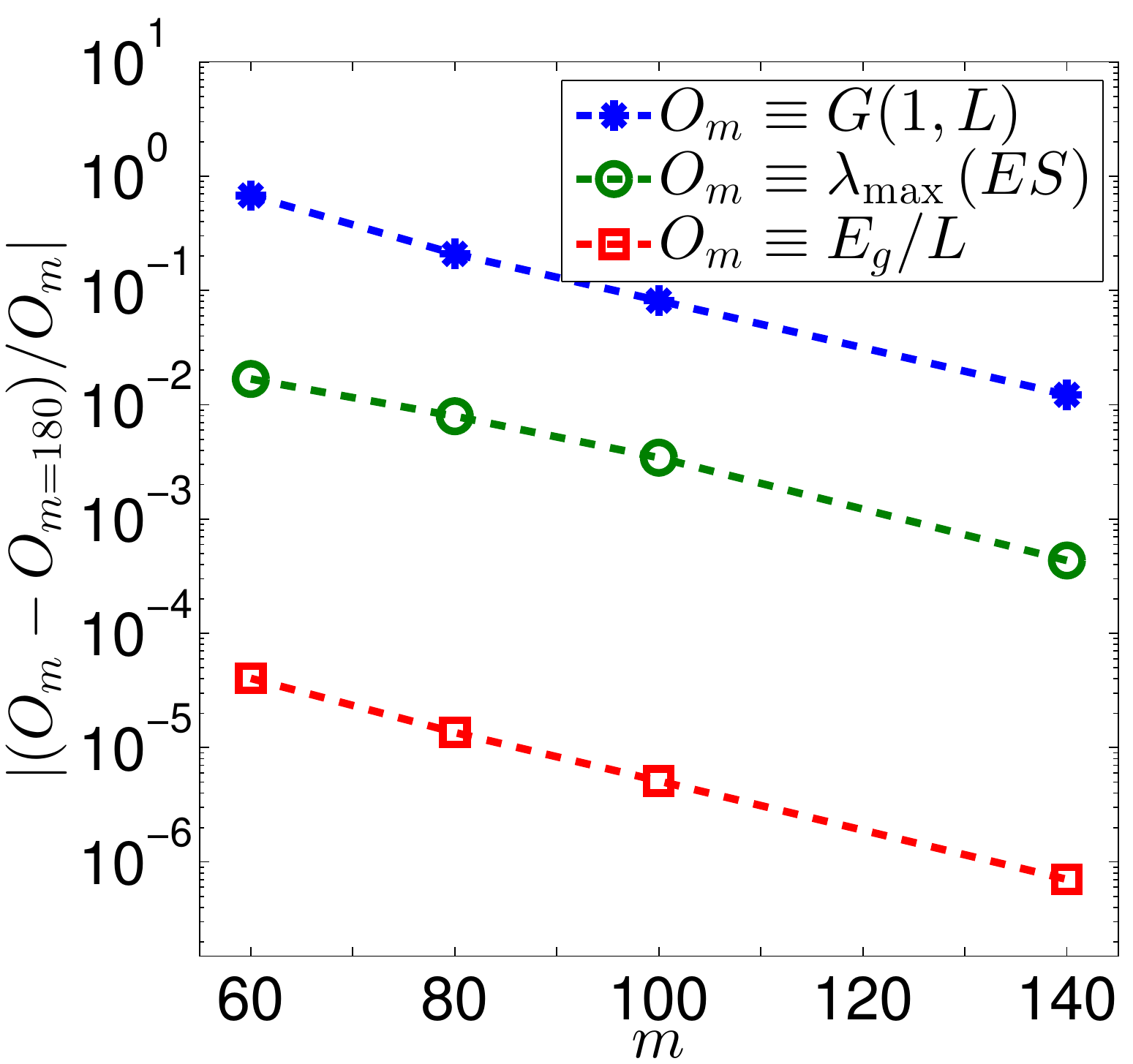}
\centering
\caption{ Numerical convergence for different observables increasing the number of kept states $m$ employed 
in the DMRG simulations, for model $\hat H$ with $L=40$ sites, at fixed parity, with filling $\nu=1/4$
 and parameters $W=-8t$, $\alpha_R=b=4t$, $U=0$. We see that all observables - edge correlations $G(1,L)$, maximum entanglement
  spectrum eigenvalue $\lambda_{\max}$, and ground state energy $e_g$ - present an exponentially
 decaying numerical error with the number of kept states $m$. }
\label{fig:SM.convergence.m}
\end{figure}

 In Fig.~\eqref{fig:SM.convergence.m} we see the convergence of some observables with 
 the number of kept states ``$m$''
  in the DMRG simulations; 
 in particular, we focus on the edge correlations $G(1,L)$, the maximum entanglement
  spectrum eigenvalue $\lambda_{\max}$, and for the ground state energy $e_g$.
   We notice that the numerical errors on the observables are small (even negligible), and 
   decay exponentially 
   with the number of kept states $m$, ensuring a high fidelity for our results.

\begin{thebibliography}{99}

\bibitem {nayak2008}C. Nayak, S. H. Simon, A. Stern, M. Freedman, and S. Das
Sarma, Rev. Mod. Phys. \textbf{80}, 1083 (2008).

\bibitem {wilczek2009}F. Wilczek, Nat. Phys. \textbf{5}, 614 (2009).

\bibitem {alicea2012}J. Alicea, Rep. Prog. Phys. \textbf{75}, 076501 (2012).

\bibitem {beenakker2013}C. W. J. Beenakker, Annu. Rev. Con. Mat. Phys. {\bf 4}, 113 (2013)

\bibitem {goldman2016} N. Goldman, J. C. Budich, and P. Zoller, Nat. Phys. {\bf 12}, 639 (2016)

\bibitem {kitaev2001} A. Kitaev,  Phys. Usp. {\bf44}, 131 (2001). 

\bibitem {Kouwenhoven}V. Mourik, K. Zou, S.M. Frolov, S.R. Plissard, E.P.A.M.
Bakkers, L.P. Kouwenhoven, Science \textbf{336}, 1003 (2012) .

\bibitem {Deng}M. T. Deng, C. L. Yu, G. Y. Huang, M. Larsson, P. Caroff, H. Q.
Xu, Nano Lett. {\bf12}, 6414 (2012).

\bibitem {Das}A. Das, Y. Ronen, Y. Most, Y. Oreg, M. Heiblum, H. Shtrikman,
Nat. Phys. {\bf8}, 887 (2012).

\bibitem {Rokhinson}L. P. Rokhinson, X. Liu, J. K. Furdyna, Nat. Phys.
\textbf{8},795 (2012).

\bibitem {nadj} S. Nadj-Perge {\it et al.}, Science {\bf 346}, 602 (2014).

\bibitem {marcus} S. M. Albrecht {\it et al.}, Nature {\bf 531}, 206 (2016).

\bibitem {liang2011}L. Jiang, T. Kitagawa, J. Alicea, A. R. Akhmerov, D.
Pekker, G. Refael, J. I. Cirac, E. Demler, M. D. Lukin, and P. Zoller, Phys.
Rev. Lett. \textbf{106}, 220402 (2011).

\bibitem{Ortiz_2014}
G. Ortiz, J. Dukelsky, E. Cobanera, C. Esebbag, and C. Beenakker, Phys. Rev. Lett.~\textbf{113}, 267002 (2014).

\bibitem{Chen_2016}
C. Chen, W. Yan, C. S. Ting, Y. Chen, and F. J. Burnell, arXiv:1701.01794 (2017).

\bibitem {cheng2011}M. Cheng and H.-H. Tu, Phys. Rev. B \textbf{84}, 094503 (2011).

\bibitem {fidkowski2011}L. Fidkowski, R. M. Lutchyn, C. Nayak and M. P. A.
Fisher, Phys. Rev. B \textbf{84}, 195436 (2011).

\bibitem {sau2011}J. D. Sau, B. I. Halperin, K. Flensberg and S. Das Sarma,
Phys. Rev. B \textbf{84}, 144509 (2011).
\bibitem{Iemini2015} F. Iemini, L. Mazza, D. Rossini, R. Fazio and S. Diehl,
Phys. Rev. Lett. \textbf{115}, 156402 (2015).

\bibitem{Lang2015} N. Lang and H. P. B\"uchler, Phys. Rev. B {\bf 92}, 041118 (2015R).

\bibitem{Kraus2013} C. V. Kraus, M. Dalmonte, M. A. Baranov, A. M. L\"auchli  and P. Zoller, Phys. Rev. Lett. \textbf{111}, 173004 (2013).


\bibitem{FootNote1}
Quasi-topological phases include a topological, gapped sector, in addition to decoupled gapless modes, see Refs.~\cite{bonderson2013,Kraus2013}.

\bibitem{cazalilla2009}M. A. Cazalilla, A. F. Ho, and M. Ueda, New J. Phys. {\bf11}, 103033 (2009).

\bibitem{gorshkov2010}A. V. Gorshkov, M. Hermele, V. Gurarie, C. Xu, P. S. Julienne, J. Ye, P. Zoller, E. Demler, M.D. Lukin, and A. M. Rey, Nat. Phys. {\bf6}, 289 (2010).

\bibitem{cazalilla2014}M. A. Cazalilla and A. M. Rey, Rep. Prog. Phys. {\bf77}, 124401 (2014).

\bibitem{stellmer2009} S. Stellmer, M.K. Tey, B. Huang, R. Grimm, and F. Schreck, Phys. Rev. Lett. {\bf103}, 200401 (2009).

\bibitem{desalvo2010} B. J. DeSalvo, M. Yan, P. G. Mickelson, Y. N. Martinez de Escobar, and T. C. Killian, Phys. Rev. Lett. {\bf105}, 030402 (2010).

\bibitem{sugawa2011} S. Sugawa, K. Inaba, S. Taie, R. Yamazaki, M. Yamashita, and Y. Takahashi, Nat. Phys. {\bf7}, 642 (2011).

\bibitem{swallows2011}M. D. Swallows, M. Bishof, Y. Lin, S. Blatt, M. J. Martin, A. M. Rey, and J. Ye, Science {\bf331}, 1043 (2011).

\bibitem{mancini2015}M. Mancini, G. Pagano, G. Cappellini, L. Livi, M. Rider, J. Catani, C. Sias, P. Zoller, M. Inguscio, M. Dalmonte, and L. Fallani, Science {\bf349}, 1510 (2015).

\bibitem{hofrichter2016}C. Hofrichter, L. Riegger, F. Scazza, M. H\"ofer, D. Rio Fernandes, I. Bloch, and S. F\:olling, Phys. Rev. X {\bf6}, 021030 (2016).

\bibitem{cappellini2014}G. Cappellini {\it et al.}, Phys. Rev. Lett. {\bf 113}, 120402 (2014).

\bibitem{scazza2014}F. Scazza, C. Hofrichter, M. H\"ofer, P. C. De Groot, I. Bloch, and S. F\"olling, Nat. Phys. {\bf10}, 779 (2014).

\bibitem{wall2016} M. L. Wall {\it et al.}, Phys. Rev. Lett. {\bf116}, 035301 (2016). 

\bibitem{livi2016} L. F. Livi {\it et al.}, Phys. Rev. Lett. {\bf117}, 220401 (2016). 

\bibitem{kolkowitz2016} S. Kolkowitz {\it et al.}, Nature {\bf 542}, 66 (2017).


\bibitem{FootNote2}
These interactions are quasi-resonant up to intermediate values of the external magnetic field.

\bibitem{Budich_2015} J. C. Budich, C. Laflamme, F. Tschirsich, S. Montangero and P. Zoller,
Phys. Rev. B \textbf{92}, 245121 (2015).

\bibitem{FootNote3}
The $b$-term is not fundamental to stabilize MQPs, but considerably simplifies a part of the theoretical analysis below.

\bibitem{Budich_2013} J. C. Budich and E. Ardonne,
Phys. Rev. B \textbf{88}, 035139 (2013).

\bibitem {White}S.~R. White, Phys. Rev. Lett. \textbf{69}, 2863 (1992).

\bibitem {Schollwoeck}U. Schollw{\"o}ck, Rev.\ Mod.\ Phys.\textbf{77}, 259 (2005).

\bibitem {gogolin_book}{A.O. Gogolin, A.A. Nersesyan, A.M. Tsvelik,
\textit{Bosonization and strongly correlated systems}, (Cambridge University
press, Cambridge, 1998).}


\bibitem {giamarchi_book}{ T. Giamarchi, \textit{Quantum Physics in one
dimension}, (Oxford University press, Oxford, 2003).}

\bibitem{MDM2017}M. Dalmonte {\it et al.}, in progress.

\bibitem {stoudenmire2011}E. M. Stoudenmire, J. Alicea, O. A. Starykh and M.
P. A. Fisher, Phys. Rev. B \textbf{84}, 014503 (2011).

\bibitem{supmat} Supplemental material, including details on the low-energy field theory.

\bibitem {Pollmann}F. Pollmann, E. Berg, A. M. Turner, M. Oshikawa, Phys. Rev.
B \textbf{81}, 064439 (2010).

\bibitem {Turner}A.~M. Turner, F. Pollmann, E. Berg, Phys. Rev. B \textbf{83},
075102 (2011).

\bibitem {tezuka2012}M. Tezuka and N. Kawakami, Phys. Rev. B \textbf{85},
140508 (R) (2012).

\bibitem{FootNote4}
This passage is reminiscent of treating the orbital degrees of freedom as a synthetic dimension~\cite{Celi2014,mancini2015,Stuhl2015}.

\bibitem {kraus2012}C. V. Kraus, S. Diehl, P. Zoller and M. A. Baranov, New J. Phys. \textbf{14}, 113036 (2012).

\bibitem{zhou2016} X. Zhou {\it et al.}, arxiv.1612.08880.


\bibitem{Celi2014} A. Celi {\it et al.}, Phys. Rev. Lett. {\bf 112}, 043001 (2014).

\bibitem{Stuhl2015}B. Stuhl {\it et al.}, Science {\bf349}, 1514 (2015)

\bibitem{bonderson2013}P. Bonderson and C. Nayak, Phys. Rev. B {\bf87}, 195451(2013).


\end{thebibliography}

\begin{thebibliography}{99}

\bibitem{SM_cheng2011}M. Cheng and H.-H. Tu, Phys. Rev. B \textbf{84}, 094503 (2011).
\bibitem{SM_Kraus2013} C. V. Kraus, M. Dalmonte, M. A. Baranov, A. M. L\"auchli  and P. Zoller, Phys. Rev. Lett. \textbf{111}, 173004 (2013).


\bibitem{SM_Iemini2015} F. Iemini, L. Mazza, D. Rossini, R. Fazio and S. Diehl,
Phys. Rev. Lett. \textbf{115}, 156402 (2015).

\bibitem{SM_Lang2015} N. Lang and H. P. B\"uchler, Phys. Rev. B {\bf 92}, 041118 (2015R).
\bibitem{SM_Budich_2015} J. C. Budich, C. Laflamme, F. Tschirsich, S. Montangero and P. Zoller,
Phys. Rev. B \textbf{92}, 245121 (2015).
\bibitem{SM_Budich_2013} J. C. Budich and E. Ardonne,
Phys. Rev. B \textbf{88}, 035139 (2013).
\end{thebibliography}
\end{document}